\documentclass[sigconf, anonymous=False]{acmart}

\AtBeginDocument{%
  }

\setcopyright{acmcopyright}
\copyrightyear{2018}
\acmYear{2018}
\acmDOI{XXXXXXX.XXXXXXX}

\acmConference[Conference acronym 'XX]{Make sure to enter the correct
  conference title from your rights confirmation email}{June 03--05,
  2018}{Woodstock, NY}
\acmPrice{15.00}
\acmISBN{978-1-4503-XXXX-X/18/06}

\usepackage{xcolor}
\usepackage{multirow}
\usepackage[linesnumbered,ruled,lined,boxed]{algorithm2e}
\usepackage{makecell}
\usepackage{siunitx}

\usepackage{amsmath}
\DeclareMathOperator*{\argmax}{argmax}

\usepackage{pifont}
\newcommand{\cmark}{\ding{51}}%

\newcommand{\markred}[1]{{\color{red}{#1}}}
\newcommand{\markblue}[1]{{\color{blue}{#1}}}

\begin{document}

\title{A Benchmarking Study of Matching Algorithms for Knowledge Graph Entity Alignment}

\author{Nhat-Minh Dao}
\affiliation{%
  \institution{École polytechnique}
  \streetaddress{Route de Saclay}
  \city{Palaiseau}
  \postcode{91128}
  \country{France}}
\email{nhat-minh.dao@polytechnique.edu}

\author{Thai V.\ Hoang}
\authornote{Corresponding author.}
\affiliation{%
  \institution{Huawei Technologies France}
  \streetaddress{18 Quai du Point du Jour}
  \city{Boulogne-Billancourt}
  \postcode{92100}
  \country{France}}
\email{thai.v.hoang@huawei.com}

\author{Zonghua Zhang}
\affiliation{%
  \institution{Huawei Technologies France}
  \streetaddress{18 Quai du Point du Jour}
  \city{Boulogne-Billancourt}
  \postcode{92100}
  \country{France}}
\email{zonghua.zhang@huawei.com}

\begin{abstract}
How to identify those equivalent entities between knowledge graphs (KGs), which is called Entity Alignment (EA), is a long-standing challenge. So far, many methods have been proposed, with recent focus on leveraging Deep Learning to solve this problem. However, we observe that most of the efforts has been paid to having better representation of entities, rather than improving entity matching from the learned representations. In fact, how to efficiently infer the entity pairs from this similarity matrix, which is essentially a matching problem, has been largely ignored by the community. Motivated by this observation, we conduct an in-depth analysis on existing algorithms that are particularly designed for solving this matching problem, and propose a novel matching method, named Bidirectional Matching (BMat). Our extensive experimental results on public datasets indicate that there is currently no single silver bullet solution for EA. In other words, different classes of entity similarity estimation may require different matching algorithms to reach the best EA results for each class. We finally conclude that using PARIS, the state-of-the-art EA approach, with BMat gives the best combination in terms of EA performance and the algorithm's time and space complexity. 
\end{abstract}



\received{20 February 2007}
\received[revised]{12 March 2009}
\received[accepted]{5 June 2009}

\maketitle

\section{Introduction}

Knowledge graph (KG) has attracted tremendous research efforts in the last decades, thanks to its capability of representing structured and interconnected real-world knowledge through entities (e.g., people, organizations, concepts), their relationships and additional textual information \cite{weikum_machine_2021}. 
As a matter of fact, the advantages of KGs have been extensively demonstrated in various downstream applications, e.g., recommendation system \cite{liu_recommender_2021}, search engine \cite{yang_search_engine_2019}, question-answering \cite{zang_question_answer_2020}). 
Nevertheless, one of the major drawbacks of real-world KGs is their incompleteness, i.e., some real-world information is not contained in KGs. 
To address this problem, link prediction, which predicts unseen links between entities in KGs \cite{boschin_combining_2022}, can be applied.  
But it cannot integrate unseen knowledge from other sources, and an Entity Alignment (EA) process has to be developed, which aims at identifying equivalent entities between different KGs and enriching their information from each other.

To date, a variety of conventional EA methods has been proposed, which are either unsupervised \cite{suchanek_paris_2011} or require a small amount of labeled data \cite{lacoste-julien_sigma_2013}. 
Their general idea is to construct initial entity similarity scores based on textual information or entity pair seeds, which is then propagated via KGs structure to align the entities. 
These solutions have been demonstrated to be efficient and robust.
In particular, PARIS \cite{suchanek_paris_2011} has remained as one of the state-of-the-art solutions that is even superior to the recently developed methods \cite{leone_critical_2022}. 
In addition to the aforementioned features, PARIS preserves simplicity and scalability, showing significant potential in real-world large-scale (e.g., millions of entities) deployment. 
Recent years have witnessed attempts to develop deep learning-based methods for entity alignment \cite{sun_benchmarking_2020}.
Generally, these methods encode structural and textual information of two KGs into a unified embedding space, and a prediction module is then utilized to infer an alignment result \cite{zhao_experimental_2022}. 
As they are mainly supervised or semi-supervised, a large portion of entity seed pairs is required to align the embedding spaces of two KGs. 
However, annotating thousands to millions of items in real-world KGs is a well-known challenge. It is time- and labor-consuming, costly, and error-prone. 

From an algorithmic perspective, an EA solution consists of two major modules: \textit{entity similarity estimation} and \textit{matching}. 
Specifically, the first module estimates pairwise similarity score between the entities of two KGs, while the second one infers the matched entity pairs from those similarity scores. 
After a careful study, we observed that most of existing EA methods focuses on entity similarity estimation, while the matching task has received much less attention than it deserves. 
For example, a majority of embedding-based EA methods does not specify any matching algorithm. They simply use ranking-based evaluation metrics, such as mean reciprocal rank or Hits@k \cite{sun_benchmarking_2020, zhao_experimental_2022}, to calculate performance.
Even though these metrics are widely used in information retrieval, they seem less suitable for the EA matching, since only Hit@1 is equivalent to the recall of a greedy matching algorithm. 

A number of work has integrated existing matching algorithms into EA solutions, such as Hungarian algorithm \cite{mao_alignment_2021}, approximation solver to optimal transport \cite{ding_viaOT_2022}, and Gale-Shapley algorithm \cite{zeng_cea_2020}. 
These algorithms have helped to get rid of many-to-1 matching and eventually improve EA performance from the greedy matching algorithm. This is a clear manifestation of the potential utility of ``better'' matching algorithms in EA.

Recently, Zeng et al.\ \cite{zeng_matching_benchmarking_2023} studied experimentally existing algorithms that can be employed for the matching module. However, their work focuses only on embedding-based similarity estimation methods and thus lacks of investigation into matching algorithms on the similarity matrices calculated from non-deep learning and entity names directly.

Without undermining the significance of entity similarity estimation methods, either conventional or embedding-based ones, we believe an insightful analysis of matching algorithms will help the community to better understand their role in and impact on developing better EA solutions. 
Specifically, our contributions are summarized as follows, 
\begin{itemize}
    \item First, we develop Bidirectional Matching (BMat), a greedy yet scalable stable matching algorithm that improves the performance of conventional EA solution.
    \item Second, we conduct a comparative study of existing matching algorithms across a wide range of datasets and entity similarity estimation methods, with an objective to recommending the most appropriate matching algorithm. 
    \item Third, we experimentally validate the advantages of PARIS algorithm over the embedding-based ones, as well as its improved performance with our new matching algorithm.
\end{itemize}

\section{Preliminaries and Related Work}
\label{sec:preliminaries}

We present first in this section the EA problem, together with a general solution pipeline for it. We then discuss existing solutions and how they are mapped to the general pipeline. 

\subsection{Knowledge graph entity alignment}

Let $E$, $R$, $A$, $V$ be the sets of entities, relations, attributes, and literals (i.e., attribute values), respectively. A KG can be defined as $G=\left(E,R,A,V,T^R,T^A\right)$ where $T^R=\{\left(h,r,t\right)\mid h,t\in E,r\in R\}$ and $T^A=\{\left(e,a,v\right)\mid e\in E, a\in A,v\in V\}$ denote the relation and attribute triples, respectively.
Given two KGs $G$ and $G'$, EA consists of finding the set $\psi$ of equivalent entities across $G$ and $G'$: 
\begin{align}
    \label{eq:entity_equivalence}
    \psi=\{(e,e') \mid e\equiv e',e\in E,e'\in E'\}
\end{align}
The equivalence operator $\equiv$ indicates that two entities refer to the same ``real-world'' object.

If there exists a set of known matched entity pairs, they can be used as \emph{seeds} in a supervised alignment setting. It is assumed that a ground-truth that includes all possible equivalences between pairs of entities is also available for performance evaluation.

\subsection{General EA framework}
\label{subsec:ea_framework}

\begin{figure}
    \centering
    \includegraphics[width=0.48\textwidth]{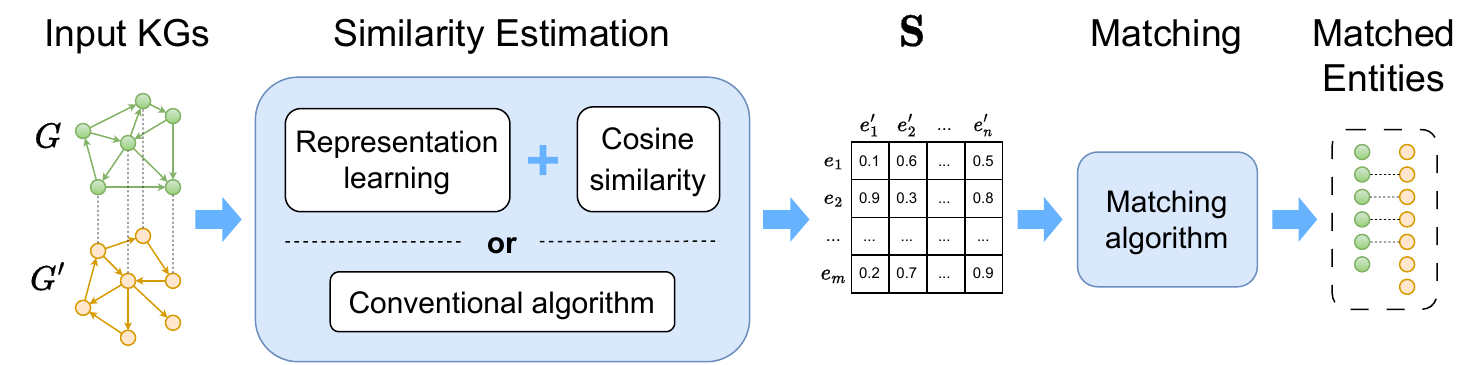}
    \caption{The general framework for existing EA solutions. The two main modules are highlighted in blue, and dash lines indicate gold links.}
    \label{fig:general_framework}
\end{figure}

Many methods have been proposed to solve the aforementioned EA problem. Their solutions are generally composed of two main functional modules, as illustrated in Figure \ref{fig:general_framework}: similarity estimation, and matching.
\emph{Similarity estimation} aims at calculating the probability of a pair of entities from two input KGs to be equivalent. Existing similarity estimation algorithms can be roughly categorized into two main groups. 
The first \cite{sun_benchmarking_2020} projects entities into a $d$-dimensional space and often uses KG data to learn their representations by optimizing an objective function. The two set of learned embedding vectors, $\mathbf{E} \in \mathbb{R}^{m\times d}$ and $\mathbf{E}' \in \mathbb{R}^{n\times d}$ with $m=|E|$ and $n=|E'|$, are then used to calculate the pairwise similarity matrix $\mathbf{S}\in \mathbb{R}^{m\times n}$ such that:
\begin{align}
    \label{eq:cosine_similarity}
    \mathbf{S}[i,j] = \mbox{cosine\_similarity}(\mathbf{E}[i,:], \mathbf{E},[j,:]).
\end{align}
The second group directly estimates $\mathbf{S}$ from the contextual data that are available in the two input KGs \cite{suchanek_paris_2011}.
Based on the estimated similarity scores of all entity pairs, the \emph{matching} module aims at solving for the entity equivalence $\psi$ (Equation \ref{eq:entity_equivalence}).

Our general framework is slightly different from the framework presented in Zeng et al.\ \cite{zeng_matching_benchmarking_2023}, which is designed for embedding-based EA only. As a result, the first module in their framework outputs the learned embeddings $\mathbf{E}, \mathbf{E}'$ and similarity calculation is a component of the second module.

\subsection{Related work}
\label{subsec:related_work}
 
Entity Alignment (EA) has emerged as an important research problem in the context of semantic webs \cite{matthews_semantic_web_2005}, along with Schema or Ontology Alignment \cite{ochieng_ontology_matching_2019}, Record Linkage \cite{barlaug_neural_2021}, etc. 
Roughly speaking, an ontology of a knowledge base is a categorization of entities, their relations, and their classes. 
While EA aims at identifying equivalent entities between a pair of knowledge bases and Schema Alignment's goal is to find the equivalent entity classes, the scope of Ontology Alignment covers all classes, relations, and entities.  
In this paper, our attention is limited to EA.  

In \cite{ferrara_auto_iden_2008}, the authors introduced the problem about identifying two instances that refer to the same object in real-life. 
The \textit{functionality} concept, which measures the popularity of relations in knowledge bases, was introduced in \cite{hogan_functionality_2010}. 
PARIS \cite{suchanek_paris_2011} then leveraged the idea to achieve robust and scalable entity alignment.
Thanks to the introduction of embedding techniques, EA problem found its new solutions by either using translational models \cite{chen_mtranse_2017, zhu_iptranse_2017, zequn_jape_2017} or Graph Convolutional Networks (GCN) \cite{wang_gcn_2018, wu_hgcn_2019, wang_gcnalign_2018, wu_rdgcn_2019}. 
Those embedding-based EA methods normally consist of entity similarity estimation and matching. 
Firstly, the entities from two KGs are encoded in a unified vector space. 
The pairwise similarity scores between cross-KG entity pairs are then obtained using cosine similarity between their embedding vectors.  
Finally, matching methods are applied to the similarity scores to obtain the alignment results. 
Most of embedding-based EA methods uses a straightforward, greedy strategy for the matching. 
However, this simple strategy suffers from many-to-1 alignment, and provides no guarantee to obtain an ``optimal'' alignment for all the entities. 
To address these issues, the authors of \cite{zeng_cea_2020} proposed to use Gale-Shapley algorithm \cite{gale_smat_1962} to solve it as a stable matching problem. 
In \cite{mao_alignment_2021}, the authors used Hungarian algorithm and optimal transport solver to deal with it as an assignment problem.
\begin{table}[t]
    \caption{The focus of existing EA benchmarking papers. Sim.\ and Mat.\ denote Similarity and Matching, respectively.}
    \label{tab:ea_survey_papers}
    \setlength{\tabcolsep}{2pt}
    \begin{tabular}{lcccccc}
        \toprule
        \multirow[c]{2}{*}{Paper} & \multirow[c]{2}{*}{Sim.} & \multirow[c]{2}{*}{Mat.} & \multirow[c]{2}{*}{PARIS} & \multirow[c]{2}{*}{Name} & \multicolumn{2}{c}{Metric} \\
         &  &  &  &  & F1-score & Hit@k \\
        \midrule
        Sun et al. \cite{sun_benchmarking_2020}            & \cmark &  & \cmark &  & \cmark & \cmark \\
        Zhang et al. \cite{zhang_industry_2020}            & \cmark &  & \cmark & \cmark & \cmark & \cmark \\
        Berrendorf et al. \cite{hiemstra_critical_2021}    & \cmark &  &  & \cmark &  & \cmark \\
        Zeng et al. \cite{zeng_comprehensive_2021}         & \cmark &  &  & \cmark &  & \cmark \\
        Zhao et al. \cite{zhao_experimental_2022}          & \cmark &  & \cmark & \cmark & \cmark & \cmark \\
        Leone et al. \cite{leone_critical_2022}            & \cmark &  & \cmark & \cmark & \cmark & \\
        Zhang et al. \cite{zhang_benchmark_2022}           & \cmark &  &  & \cmark &  & \cmark \\
        Fanourakis et al. \cite{fanourakis_knowledge_2022} & \cmark &  &  & \cmark &  & \cmark \\
        Zeng et al. \cite{zeng_matching_benchmarking_2023} &  & \cmark &  & \cmark & \cmark & Hit@1 \\
        Our work                                           &  & \cmark & \cmark & \cmark & \cmark & Hit@1 \\
        \bottomrule
    \end{tabular}
\end{table}


To comparatively analyze the existing EA solutions, several benchmarking studies have been conducted. They are summarized in Table \ref{tab:ea_survey_papers} with a highlight on their scope. 
For example, in \cite{sun_benchmarking_2020}, Sun et al. conducted benchmarking on embedding-based EA solutions, and concluded that incorporating textual information and bootstrapping strategy into EA solution can achieve better performance. 
Moreover, using CSLS \cite{lample_word_2018} as a stable matching algorithm with embedding-based methods can improve performance. 
In \cite{leone_critical_2022}, the authors reported a comprehensive survey that covers both conventional (e.g., PARIS) and embedding-based methods. 
One of the interesting conclusions is that PARIS and its variant PARIS+ achieved superior performance than the embedding-based methods. 
In \cite{zeng_matching_benchmarking_2023}, the authors focused more on a benchmarking of matching algorithms for embedding-based EA solutions. 
One of the key findings, for instance, is that the classical matching algorithm (i.e, Hungarian algorithm) can outperform the algorithms that are tailored to KGs under 1-to-1 alignment constraint.

To complement with the previous works, our benchmarking study is intended to cover a wide range of matching methods that are used for both \textit{conventional} (non-deep learning) and \textit{embedding-based} EA solutions. 
In particular, entity similarity estimation methods based on entity names are included as well.   
It is also worth noting that the study always takes into account the general assumption about EA, i.e., 1-to-1 alignment or no duplicate entities in KGs.

\section{Matching Algorithms}
\label{sec:matching_algorithms}

In this section, we present a set of algorithms for the matching module introduced in Section \ref{subsec:ea_framework}. 
These algorithms are originally designed for two different problems: 
\emph{linear assignment} and \emph{stable matching}.
We carefully select the most representative ones for our experiments reported in Section \ref{sec:experiments}, which are DInf, Hun. Sink-o, Sink-d for linear assignment and SMat, BMat for stable matching.
A summary of the key features of matching algorithms is listed in Table \ref{tab:algo_characteristics}.

Without loss of generality, we assume that $|E| = |E'| = n$ in the following discussions about time and space complexity. 

\begin{table}[t]
    \caption{Comparison of matching algorithms.}
    \label{tab:algo_characteristics}
    \begin{tabular}{lcccc}
        \toprule
        & \multirow[c]{2}{*}{\makecell{1-to-1\\ matching}} & \multirow[c]{2}{*}{\#Output pairs} & \multicolumn{2}{c}{Complexity} \\ 
        &  &  & Time & Space  \\ 
        \midrule
        Dinf   &            & $|E|$ or $|E'|$  & $\mathcal{O}(n^2)$       & $\mathcal{O}(n^2)$ \\
        Hun.   & \checkmark & $\min(|E|,|E'|)$ & $\mathcal{O}(n^3)$       & $\mathcal{O}(n^2)$ \\
        Sink-o &            & $|E|$ or $|E'|$  & $\mathcal{O}(n^2k)$      & $\mathcal{O}(n^2)$ \\
        Sink-d &            & $|E|$ or $|E'|$  & $\mathcal{O}(n^3\log n)$ & $\mathcal{O}(n^2)$ \\
        SMat   & \checkmark & $\min(|E|,|E'|)$ & $\mathcal{O}(n^2\log n)$ & $\mathcal{O}(n^2)$ \\
        BMat   & \checkmark & $\min(|E|,|E'|)$ & $\mathcal{O}(n^2)$       & $\mathcal{O}(n^2)$ \\
        \bottomrule
    \end{tabular}
\end{table}

\subsection{Linear assignment}

The task of matching entities according to their pairwise similarity scores can be formulated as an assignment problem, which is a fundamental and well-studied problem in combinatorial optimization. 
In its linear form, the objective is to maximize the sum of pairwise similarity scores of matched entity pairs, subject to the 1-to-1 constraint. The linear assignment problem can be written as:
\begin{equation}
    \label{eq:linear_assignment}
    \mathbf{P}^* = \argmax_{\mathbf{P} \in \mathbb{P}_n} \, \langle \mathbf{P}, \mathbf {S} \rangle _{\mathrm {F} },
\end{equation}
where $\mathbf{P}$ is an instance of the set of all $n$-dimensional permutation matrices $\mathbb{P}_n$, and $\langle \cdot \rangle _{\mathrm {F}}$ represents the Frobenius inner product. $\mathbf{P}^*[i,j] = 1$ indicates that $(e_i, e'_j)$ is a matched entity pair.

\subsubsection{Greedy algorithm (DInf)}

DInf gives an approximate solution to the problem in Equation \ref{eq:linear_assignment}. 
It scans $\mathbf{S}$ only once and performs the matching row-wise or column-wise, depending on the matching direction. 
For row-wise, the approximate solution is defined as follows,
\begin{equation}
    \label{eq:algo_greedy}
    \mathbf{\hat{\mathbf{P}}}[i,j] = 1 \Leftrightarrow j = \argmax \mathbf{S}[i, :].
\end{equation}
DInf belongs to the rank-based class of algorithms that is often used to calculate the Hit@k metrics. In fact, since DInf's solution is the same as Top-1 ranking, the F1-score calculated directly from DInf's output is equal to Hit@1.
DInf has $\mathcal{O}(n^2)$ time and $\mathcal{O}(n^2)$ space complexity. It does not guarantee 1-to-1 matching.

\subsubsection{Hungarian algorithm (Hun.)}

Hun.\ is a popular combinatorial optimization algorithm to globally solve the assignment problem in Equation \ref{eq:linear_assignment} in polynomial time \cite{jonker_shortest_1987}. It was used in previous work for the matching problem in EA \cite{xu_coordinated_2020, mao_alignment_2021}.
Hun.\ has $\mathcal{O}(n^3)$ time and $\mathcal{O}(n^2)$ space complexity. Its solution respects the 1-to-1 matching constraint. 

\subsection{Optimal transport}

The linear assignment problem could be seen as a special case of the optimal transport problem (OT). 
OT is a mathematical optimization problem that deals with the optimal transportation of resources from one location to another. 
While many algorithms have been developed to solve OT, we present here two specific algorithms that have been used for the matching problem of EA.


\subsubsection{Sinkhorn operator (Sink-o)}

Sink-o is a simple analog of the softmax operator for matrix and is defined as 
\begin{align*}
    F^0(\mathbf{S}) &= \exp(\mathbf{S}) \\
    F^k(\mathbf{S}) &= \mathcal{N}_c(\mathcal{N}_r(F^{k-1}(\mathbf{S}))) \\
    \mbox{Sinkhorn}(\mathbf{S}) &= \lim_{k \rightarrow \infty} F^k(\mathbf{S})
\end{align*}
where $\mathcal{N}_c$ and $\mathcal{N}_r$ are the column-wise and row-wise normalization operators. Mena et al.\ \cite{mena_sinkhorn_approx_2018} showed that we can approximate $\mathbf{P}^* \approx \mbox{Sinkhorn} (\mathbf{S}/\tau)$ with a small value of $\tau$. 
Based on this observation, Sink-o was used for the matching problem in EA \cite{mao_alignment_2021} by applying DInf to the approximated permutation matrix $\hat{\mathbf{P}}$ output by Sink-o. 
It has $\mathcal{O}(n^2 k)$ time and $\mathcal{O}(n^2)$ space complexity.

\subsubsection{Sinkhorn distance (Sink-d)}

Sink-d provides a way to measure the dissimilarity between two probability distributions \cite{cuturi_sinkhorn_distance_2013}. 
It is a distance metric that can be applied to solve various optimization problems, including the assignment problem. 
By adding entropy as regularization to OT, Sink-d can stabilize the optimization process and avoid degenerate solutions. 
This strategy also allows for efficient computation, especially for large-scale problems, as the Sinkhorn-Knopp algorithm has a fast convergence rate. 

Sink-d was used for the matching problem in EA \cite{gao_clusterea_2022, ge_make_2021} by applying DInf to the approximated permutation matrix $\hat{\mathbf{P}}$ output by Sink-d. 
It has $\mathcal{O}(n^3 \log n)$ time and $\mathcal{O}(n^2)$ space complexity.

\subsection{Stable matching}

Entity matching can be also formulated as a stable marriage problem, which finds a stable matching between two equally sized sets of entities given an ordering of preferences for each entity. 
A matching is \emph{stable} if there does not exist any pair $(e_i, e'_j)$ which both prefer each other to their current matched entity under the matching.
There exists a lattice of solutions to this problem \cite{hwang_lattice_1982}. 
We present here two algorithms that can be used to construct the three special elements of this lattice: top, bottom, and center.

\subsubsection{Gale-Shapley algorithm (SMat)}

SMat \cite{gale_smat_1962} is a popular algorithm to solve the stable matching problem. It can be used to construct the top and bottom elements of the stable matching lattice. 
SMat operates by iteratively proposing and accepting/rejecting matches. It begins with an unmatched entity $e \in E$ and propose to match it to an entity $e' \in E'$ based on its preference order. $e'$ can accept to be matched to $e$ if it is not yet matched or its current match is less-preferred than $e$. 

SMat was used recently for the matching problem in EA \cite{zeng_cea_2020, zhu_raga_2021}. It has $\mathcal{O}(n^2)$ time and $\mathcal{O}(n^2)$ space complexity. However, it has an additional overhead of generating preference orders from $\mathbf{S}$, which has $\mathcal{O} (n^2 \log n)$ time complexity due to the sorting operations. SMat's solution respects the 1-to-1 matching constraint. 

\subsubsection{Bidirectional matching (BMat)}

BMat is an algorithm that can construct the center element of the stable matching lattice. 
It runs on $\mathbf{S}$ and does not need to generate preference orders beforehand. In each iteration, BMat matches an unmatched entity $e_i \in E$ to an unmatched entity $e'_j \in E'$ if and only if $j = \argmax \mathbf{S}[i, :]$ and $i = \argmax \mathbf{S}[:, j]$. The full listing of BMat is given in Algorithm \ref{alg:bidirectional_matching}.
\begin{algorithm}[t]
    \caption{Bidirectional matching algorithm}
    \label{alg:bidirectional_matching}
    \DontPrintSemicolon
    \KwData{Pairwise similarity matrix $\mathbf{S} \in \mathbb{R}^{|E| \times |E'|}$}
    \KwResult{Set of equivalent pairs $\psi$}
    $\psi \gets \emptyset$ \;
    $R \gets \{0, 1, \cdots, |E|-1\}$  \;
    \While{$R$}{
        $M \gets \emptyset$ \;
        \For{$r \in R$}{
            $c_m \gets \argmax \mathbf{S}[r, :]$ \;
            $r_m \gets \argmax \mathbf{S}[:, c_m]$ \;
            \If{$r = r_m$}{
                $\mathbf{S}[r_m, :] \gets 0$ \;
                $\mathbf{S}[:, c_m] \gets 0$ \;
                $\psi \gets \psi + (r_m, c_m)$ \;
                $M \gets M + r_m$ \;
            }
        }
        $R \gets R - M$ \;
    }
\end{algorithm}

To the best of our knowledge, BMat has not yet been used for the matching problem in EA. 
A simplified version was employed to find ``trustworthy'' alignment in a continual EA context \cite{sattler_facing_2022}. 
Its closely-related counterpart for the maximum weighted matching problem in graphs was proposed by Preis \cite{preis_approx_1999}. 

BMat has $\mathcal{O}(n^2)$ time and $\mathcal{O}(n^2)$ space complexity. Its solution respects the 1-to-1 matching constraint.  

\subsection{Matching with a threshold}
\label{subsec:matching_with_threshold}

The aforementioned matching algorithms do not pay attention to the similarity score of each matched entity pair, but to the global sum of the scores or the stability of the matching. 
As a consequence, some matched entity pairs might have small similarity score, making them less pertinent. 
We use a threshold $\theta$ to refine $\psi$, the set of matched entity pairs, as follows, 
\begin{align}
    \label{eq:matching_with_threshold}
    \hat\psi = \{(e_i, e'_j) \mid (e_i, e'_j) \in \psi, \mathbf{S}[i,j] \ge \theta \}
\end{align}
Besides removing less-probable entity pairs, this strategy can also be used to handle ``dangling'' entities \cite{sun_dangling_2021}. 
Without it, the number of matched entity pairs output by each matching algorithm are listed in Table \ref{tab:algo_characteristics}. 
We set $\theta = 0.5$ in our experiments.

\section{Experiments}
\label{sec:experiments}

In this section, we benchmark the performance of matching algorithms discussed in Section \ref{sec:matching_algorithms} on common EA public datasets.
Some algorithms, such as CSLS \cite{lample_word_2018} and RInf \cite{zeng_entity_2022} (designed for improving the pairwise similarity matrix $\mathbf{S}$) and RL \cite{zeng_reinforcement_2021} (which aims at identifying entity equivalence) are not included, since they have been shown to have inferior performance on benchmarking datasets in a previous study \cite{zeng_matching_benchmarking_2023}. 

\subsection{Experimental settings}
\label{subsec:exp_setting}

As a follow-up of the previous benchmarking on matching algorithms for KG entity alignment \cite{zeng_matching_benchmarking_2023}, we adopted similar settings by using same public datasets, same evaluation and similarity metrics.

\paragraph{Datasets:} \textbf{DBP15K} dataset \cite{damato_cross-lingual_2017} contains three cross-lingual KG pairs extracted from DBpedia \cite{hutchison_dbpedia_2007} (version 2016-04): French-English (FR\_EN), Japanese-English (JA\_EN), and Chinese-English (ZH\_EN). Originally, each KG pair has 15000 gold links (i.e., pairs of equivalent entities) in its ground-truth. However, there exist a non-negligible number of equivalent entity pairs that are not in the ground-truth. Wang et al. \cite{sattler_facing_2022} recently employed interlanguage links in DBpedia and identified 3967, 3969, 3469 additional pairs for FR\_EN, JA\_EN, ZH\_EN, respectively. The updated ground-truth, which is composed of the original pairs and the newly identified pairs, is used in our study.
\textbf{SRPRS} \cite{guo_learning_2019} contains KG pairs extracted from DBpedia, Wikipedia \cite{vrandecic_wikidata_2014} and YAGO \cite{suchanek_yago_2007}. We use four KG pairs, each contains 15000 gold links, from the V1 version: two mono-lingual (DBpedia-Wikidata (DBP\_WD), DBpedia-YAGO (DBP\_YG)) and two cross-lingual (English-French (EN\_FR), English-German (EN\_DE)). KGs of the V1 version are relatively sparse, with a large amount of entities having low degrees to reflect real-world KGs.
\textbf{DWY100K} is a larger dataset containing two mono-lingual KG pairs: DBpedia-Wikipedia (DBP\_WD) and DBpedia-YAGO (DBP\_YG). We mainly use this dataset to evaluate the scalability of matching algorithms since each KG pair has 100000 gold links. 
Statistics of the three datasets are listed in Table \ref{tab:dataset_statistics}. For each KG pair, the number of gold links, the bijectivity of their entity sets and the existence of entity names are further given in Table \ref{tab:dataset_characteristics}.

\begin{table}
    \caption{The number of entities ($|E|$), literals ($|L|$), relations ($|R|$), attributes ($|A|$), and the average number of relations and attributes per entity ($|T^A|/|E|$ and $|T^A|/|E|$) for each KG pair of the three public datasets.}
    \label{tab:dataset_statistics}
    \setlength{\tabcolsep}{3pt}
    \begin{tabular}{llrrrrcS[table-format=2.2]}
        \toprule
         &  & \multicolumn{1}{c}{$|E|$} & \multicolumn{1}{c}{$|L|$} & \multicolumn{1}{c}{$|R|$} & \multicolumn{1}{c}{$|A|$} & $|T^R|/|E|$ & \multicolumn{1}{c}{$|T^A|/|E|$} \\
        \midrule
        \multirow[c]{6}{*}{\rotatebox[origin=c]{90}{DBP15K}} 
         & \multirow[c]{2}{*}{FR\_EN} 
            & 19661 & 214202 &  903 & 4547 & 5.39 & 26.89 \\
         &  & 19993 & 189149 & 1208 & 6422 & 5.79 & 28.84 \\
        \cline{2-8}
         & \multirow[c]{2}{*}{JA\_EN} 
            & 19814 & 138502 & 1299 & 5881 & 3.90 & 17.89 \\
         &  & 19780 & 181937 & 1153 & 6066 & 4.73 & 25.14 \\
        \cline{2-8}
         & \multirow[c]{2}{*}{ZH\_EN} 
            & 19388 & 171310 & 1701 & 8113 & 3.63 & 19.58 \\
         &  & 19572 & 197254 & 1323 & 7173 & 4.86 & 29.01 \\
        \cline{1-8}
        \multirow[c]{8}{*}{\rotatebox[origin=c]{90}{SRPRS}} 
         & \multirow[c]{2}{*}{EN\_FR} 
            & 15000 & 38141 & 221 & 296 & 2.43 & 4.72 \\
         &  & 15000 & 37414 & 177 & 415 & 2.24 & 3.76 \\
        \cline{2-8}
         & \multirow[c]{2}{*}{EN\_DE} 
            & 15000 & 33403 & 222 & 296 & 2.56 & 4.18 \\
         &  & 15000 & 39021 & 120 & 193 & 2.49 & 9.50 \\
        \cline{2-8}
         & \multirow[c]{2}{*}{DBP\_WD} 
            & 15000 &  39739 & 253 & 363 & 2.56 & 4.80 \\
         &  & 15000 & 123613 & 144 & 652 & 2.68 & 9.09 \\
        \cline{2-8}
         & \multirow[c]{2}{*}{DBP\_YG}
            & 15000 & 38598 & 223 & 320 & 2.25 & 4.62 \\
         &  & 15000 & 18543 &  30 &  22 & 2.44 & 1.50 \\
        \cline{1-8}
        \multirow[c]{2}{*}{\rotatebox[origin=c]{90}{DWY100K}} 
         & \multirow[c]{2}{*}{DBP\_WD}
            & 100000 & 154002 & 330 & 351 & 4.63 & 3.86 \\
         &  & 100000 & 656437 & 220 & 729 & 4.49 & 7.90 \\
        \cline{2-8}
         & \multirow[c]{2}{*}{DBP\_YG}
            & 100000 & 169985 & 302 & 334 & 4.29 & 4.52 \\
         &  & 100000 &  78191 &  31 &  23 & 5.03 & 1.18 \\
        \bottomrule
    \end{tabular}
\end{table}

\begin{table*}
    \caption{Some characteristics of KG pairs (\#gold links, bijectivity, cross-lingual or mono-lingual, the existence of name) and pairwise similarity matrices (size, shape and sparsity in case of PARIS).}
    \label{tab:dataset_characteristics}
    \setlength{\tabcolsep}{3.1pt}
    \begin{tabular}{llccccccc}
        \toprule
         &  & \multirow[c]{2}{*}{\#Gold links} & \multirow[c]{2}{*}{Bijective} & \multirow[c]{2}{*}{Cross-lingual} & \multirow[c]{2}{*}{Name} & \multicolumn{2}{c}{Shape and size (GB) of $\mathbf{S}$} & \multirow[c]{2}{*}{\makecell{PARIS: size (GB) \\ and sparsity (\%)}} \\
         &  &  &  &  &  & PARIS, name-based & GCN-Align, RREA &  \\
        \midrule
        \multirow[c]{3}{*}{DBP15K} 
         & FR\_EN  & 18967  &        & \cmark & \cmark & (19661, 19993), 2.929    & (10500, 10500), 0.821  & 0.043, 1.48\\
         & JA\_EN  & 18969  &        & \cmark & \cmark & (19814, 19780), 2.920    & (10500, 10500), 0.821  & 0.040, 1.38\\
         & ZH\_EN  & 18469  &        & \cmark & \cmark & (19388, 19572), 2.827    & (10500, 10500), 0.821  & 0.066, 2.34\\
        \hline
        \multirow[c]{4}{*}{SRPRS} 
         & EN\_FR  & 15000  & \cmark & \cmark & \cmark & (15000, 15000), 1.676    & (10500, 10500), 0.821  & 0.015, 0.88\\
         & EN\_DE  & 15000  & \cmark & \cmark & \cmark & (15000, 15000), 1.676    & (10500, 10500), 0.821  & 0.022, 1.29\\
         & DBP\_WD & 15000  & \cmark &        &        & (15000, 15000), 1.676    & (10500, 10500), 0.821  & 0.020, 1.20\\
         & DBP\_YG & 15000  & \cmark &        & \cmark & (15000, 15000), 1.676    & (10500, 10500), 0.821  & 0.011, 0.68\\
        \hline
        \multirow[c]{2}{*}{DWY100K} 
         & DBP\_WD & 100000 & \cmark &        &        & (100000, 100000), 74.506 & (70000, 70000), 36.508 & 0.205, 0.28\\
         & DBP\_YG & 100000 & \cmark &        & \cmark & (100000, 100000), 74.506 & (70000, 70000), 36.508 & 0.376, 0.51\\
        \bottomrule
    \end{tabular}
\end{table*}

\paragraph{Evaluation metric:} We use precision, recall, F1-score as our evaluation metrics. 
All values reported in our work are in percentage scale, ranging from 0 to 100. A perfect matching should have F1-score = 100.

\paragraph{Similarity metric:} For embedding-based approaches, we use cosine similarity to compute a pairwise similarity matrix $\mathbf{S}$ from the two sets of embedding vectors $\mathbf{E}$ and $\mathbf{E}'$ that correspond to the two sets of entities in the two input KGs $G$ and $G'$, respectively.

\subsection{Similarity matrix estimation approaches}
\label{subsection:matrix_estimation}

The general EA framework described in Section \ref{subsec:ea_framework} is composed of two modules. The first one aims at estimating the pairwise similarity of entity pairs. It takes two KGs as input and outputs a similarity matrix. In this work, we employ PARIS \cite{suchanek_paris_2011}, GCN-Align\cite{wang_gcnalign_2018}, RREA \cite{mao_relational_2020}, and name-based similarity estimation for the first module. 

\emph{PARIS} is a conventional unsupervised method that has been shown to have very competitive performance in previous EA benchmarking on the first module \cite{sun_benchmarking_2020, zhang_industry_2020, zhao_experimental_2022, leone_critical_2022}. It does not learn a representation for each entity, but directly estimates the similarity of each entity pair. In its original form, PARIS keeps similarity scores of highly-probable matched pairs only and does not output a similarity matrix. Although this unique mechanism allows it to work with large-scale KGs, some adaptations are required so that more sophisticate matching algorithms run.
For benchmarking purpose, we output the estimated similarity scores of highly-probable matched pairs of the last iteration as a sparse matrix. 

\emph{GCN-Align} and \emph{RREA} are embedding-based supervised methods. They learn functions to map each entity into an embedding space using the structural information of input KGs. They are representative methods of their class and were used in the previous benchmarking on matching algorithms \cite{zeng_matching_benchmarking_2023}. The GCN-Align and RREA embeddings used in our study are from that benchmarking to facilitate the comparison. 

\emph{Name-based similarity estimation:} Entity names have been shown useful for EA in previous works \cite{zhao_experimental_2022}. In this study, we calculate three pairwise similarity matrices only from entity names for KG pairs that have this auxiliary information (Table \ref{tab:dataset_characteristics}). 
The first matrix is calculated using Sorensen-Dice coefficient \cite{dice_measures_1945, sorenson_method_1948}. For each entity pair $(e_i,e'_j)$, $\mathbf{S}[i,j]={\frac {2|X\cap Y|}{|X|+|Y|}}$ where $X$ and $Y$ are the sets of tokens extracted from their names, respectively. For non-English KGs, we use Google Translate to convert their entity names into English. 
The second matrix is computed from the fastText embeddings \cite{joulin_fasttext_2016} of translated entity names, which are available from the previous benchmarking \cite{zeng_matching_benchmarking_2023}. 
The third matrix is calculated from the LaBSE embeddings \cite{feng_LaBSE_2022} of original entity names, without translation. 

Among the aforementioned approaches to estimate entity similarity, PARIS and name-based ones are unsupervised while GCN-Align and RREA are supervised. GCN-Align and RREA thus require \emph{seeds}, i.e. known matched entity pairs, in their settings.

\subsection{Similarity matrix characteristics}
\label{subsection:matrix_characteristics}

\emph{Shape and size}: As PARIS and name-based methods are unsupervised, they use 100\% of input entities to compute $\mathbf{S}$. The shape of $\mathbf{S}$ in this case is $(|E|, |E'|)$. RREA and GCN-Align, on the other hand, use 20\%, 10\%, and 70\% splits for training, validation, and testing. The shape of $\mathbf{S}$ in the testing phase is only $(0.7 \times |\psi|, 0.7 \times |\psi|)$, where $\psi$ is the set of gold links (Equation \ref{eq:entity_equivalence}).
Table \ref{tab:dataset_characteristics} lists the shape and size of $\mathbf{S}$ for all KG pairs for unsupervised (PARIS, name-based) and supervised (GCN-Align, RREA) methods. The size values are calculated assuming that the matrices are dense with elements are stored in \texttt{float64}.
Since the size of $\mathbf{S}$ is directly proportional to $|E| \times |E'|$, it increases rapidly from 1.676 GB for SRPRS to 74.506 GB for DWY100K in unsupervised settings. In supervised settings, the size of $\mathbf{S}$ is roughly half that number, which is 36.508 GB for DWY100K.
For small-sized datasets such as DBP15K or SRPRS, existing matching algorithms can run on commodity servers. For medium-sized ones such as DWY100K, dedicated servers with high memory space are required. And for large-scale ones that have millions of entities \cite{ge_largeea_2021}, memory allocation for a dense $\mathbf{S}$ is simply impossible without special hardware systems. 
Consequently, similarity estimation approaches that output a dense similarity matrix for the matching step (e.g., GCN-Align, RREA, and name-based) do not scale due to the size of $\mathbf{S}$. One possible exception is DInf, of which an efficient implementation can run directly on the sets of embedding vectors.

\emph{Sparsity}: PARIS has a mechanism to suppress the similarity of less-probable entity pairs to zero, leading to a sparse $\mathbf{S}$. Table \ref{tab:dataset_characteristics} lists the size of memory occupied by and the percentage of non-zero elements in this matrix for difference KG pairs. It can be seen that $\mathbf{S}$ output by PARIS is highly sparse, requiring on-average only 1.12 \% of eligible space for non-zero elements. For example, the size of $\mathbf{S}$ for the two KG pairs DBP\_WD and DBP\_YG of DWY100K are only 0.205 GB and 0.376 GB, respectively.
Thanks to this sparsity, PARIS scales and can probably run on large-scale datasets.

\subsection{Adaptation of matching algorithms}

\emph{Non-square matrix:} When $|E| \neq |E'|$ such as KG pairs of DBP15K dataset, $\mathbf{S}$ output from PARIS or name-based methods is non-square. All matching algorithms discussed in Section \ref{sec:matching_algorithms}, except SMat, accept a non-square matrix as input. For SMat, we adapt by introducing dummy entities to the KG that has a fewer number of entities. This can be done effectively by zero-padding rows or columns to $\mathbf{S}$ in order to make it square. All matched entity pairs output by SMat that involve one of these dummy entities are discarded.

\emph{Sparse matrix} output by PARIS can be converted to dense matrix before inputting it to matching algorithms. However, in order to benefit from its space efficiency, we adapt some matching algorithms (DInf, Sink-o, SMat, BMat) so that they can directly work with sparse matrix. More specifically, we only convert a row or column of $\mathbf{S}$ from sparse to dense format on-the-fly when it is necessary, i.e.\  when a full row or column is needed as input to an $\argmax$ operator. For Hun. and Sink-d, we rely on implementations from SciPy and FML\footnote{FML library: \href{https://github.com/fwilliams/fml}{https://github.com/fwilliams/fml}} libraries that only accept dense matrix as input.

\subsection{Overall EA performance}

\begin{table*}[t]
    \caption{Performance of matching algorithms based on PARIS, GCN-Align, RREA for DBP15K and SRPRS datasets.}
    \label{tab:performance_prf1_05}
    \setlength{\tabcolsep}{2.7pt}
    \begin{tabular}{|ll|ccc|ccc|ccc|ccc|ccc|ccc|ccc|}
        \toprule
         &  & \multicolumn{9}{c|}{DBP15K} & \multicolumn{12}{c|}{SRPRS} \\
         \cline{3-23}
         &  & \multicolumn{3}{c|}{FR\_EN} & \multicolumn{3}{c|}{JA\_EN} & \multicolumn{3}{c|}{ZH\_EN} & \multicolumn{3}{c|}{EN\_FR} & \multicolumn{3}{c|}{EN\_DE} & \multicolumn{3}{c|}{DBP\_WD} & \multicolumn{3}{c|}{DBP\_YG} \\
         &  & P & R & F1 & P & R & F1 & P & R & F1 & P & R & F1 & P & R & F1 & P & R & F1 & P & R & F1 \\
        \midrule
        \multirow[c]{7}{*}{\rotatebox[origin=c]{90}{PARIS}} & Native & \textbf{98.6} & 80.1 & 88.4 & \textbf{97.8} & 75.4 & 85.2 & \textbf{97.8} & 78.0 & 86.8 & \textbf{98.7} & 69.2 & 81.3 & \textbf{98.7} & 85.3 & 91.5 & \textbf{98.9} & 70.2 & 82.1 & \textbf{94.1} & 23.9 & 38.1 \\
         \cline{3-23}
         & DInf & 86.8 & 80.1 & 83.4 & 87.2 & 75.9 & 81.2 & 89.2 & 78.3 & 83.4 & 77.4 & 69.3 & 73.1 & 93.8 & 85.7 & 89.6 & 85.4 & 70.5 & 77.2 & 55.4 & 24.2 & 33.7 \\
         & Hun. & 87.3 & 78.1 & 82.4 & 89.4 & 75.4 & 81.8 & 90.1 & 77.7 & 83.5 & 94.2 & 71.1 & 81.0 & 96.4 & 86.3 & 91.1 & 95.6 & 71.2 & 81.6 & 80.7 & 25.1 & 38.3 \\
         & Sink-o & 77.0 & 71.0 & 73.9 & 80.5 & 70.1 & 75.0 & 80.3 & 70.5 & 75.1 & 76.7 & 68.7 & 72.5 & 91.8 & 83.9 & 87.7 & 84.6 & 69.7 & 76.4 & 56.8 & 24.8 & 34.5 \\
         & Sink-d & 82.9 & 76.1 & 79.4 & 84.5 & 73.4 & 78.5 & 86.3 & 75.6 & 80.6 & 84.4 & 71.1 & 77.2 & 93.8 & 85.6 & 89.5 & 88.5 & 71.0 & 78.8 & 70.8 & 25.0 & 37.0 \\
         & SMat & 96.8 & \underline{85.3} & \underline{90.7} & 95.7 & \underline{79.4} & \underline{86.8} & 96.4 & \underline{82.0} & \underline{88.6} & 97.3 & \underline{72.7} & \underline{83.2} & 98.1 & \textbf{87.5} & \textbf{92.5} & 98.0 & \underline{72.2} & \underline{83.1} & 84.7 & \underline{25.8} & \underline{39.6} \\
         & BMat & 96.8 & \textbf{85.3} & \textbf{90.7} & 95.7 & \textbf{79.4} & \textbf{86.8} & 96.4 & \textbf{82.0} & \textbf{88.6} & 97.3 & \textbf{72.7} & \textbf{83.2} & 98.0 & 87.4 & 92.4 & 98.0 & \textbf{72.2} & \textbf{83.1} & 84.7 & \textbf{25.8} & \textbf{39.6}\\
         \hline
        \multirow[c]{6}{*}{\rotatebox[origin=c]{90}{GCN-Align}} & DInf & 28.6 & 28.6 & 28.6 & 29.6 & 29.6 & 29.6 & 29.2 & 29.2 & 29.2 & 17.0 & 17.0 & 17.0 & 32.4 & 32.2 & 32.3 & 20.2 & 20.2 & 20.2 & 25.3 & 25.3 & 25.3 \\
         & Hun. & \textbf{48.6} & \textbf{48.5} & \textbf{48.5} & \textbf{48.3} & \textbf{48.0} & \textbf{48.1} & \textbf{45.2} & \textbf{45.0} & \textbf{45.1} & \underline{25.0} & 24.7 & 24.8 & \underline{39.1} & 38.5 & 38.8 & 28.8 & 28.5 & 28.6 & \textbf{33.3} & 33.0 & \underline{33.2} \\
         & Sink-o & 19.7 & 19.6 & 19.6 & 22.2 & 22.0 & 22.1 & 20.5 & 20.3 & 20.4 & 14.6 & 14.4 & 14.5 & 21.5 & 20.8 & 21.2 & 17.3 & 17.0 & 17.2 & 18.8 & 18.7 & 18.7 \\
         & Sink-d & 48.5 & 48.4 & 48.4 & 47.9 & 47.8 & 47.9 & 44.9 & 44.7 & 44.8 & \textbf{25.0} & \textbf{24.8} & \textbf{24.9} & \textbf{39.1} & \textbf{38.7} & \textbf{38.9} & \textbf{29.2} & \textbf{28.9} & \textbf{29.1} & 33.2 & \textbf{33.1} & \textbf{33.2} \\
         & SMat & 39.3 & 38.8 & 39.1 & 42.8 & 41.9 & 42.3 & 39.1 & 38.2 & 38.6 & 23.8 & 23.1 & 23.5 & 38.1 & 37.0 & 37.5 & 26.8 & 26.0 & 26.4 & 31.8 & 31.2 & 31.5 \\
         & BMat & 39.3 & 38.8 & 39.1 & 42.8 & 41.9 & 42.3 & 39.1 & 38.2 & 38.6 & 23.8 & 23.1 & 23.5 & 38.1 & 37.1 & 37.6 & 26.8 & 26.0 & 26.4 & 31.8 & 31.2 & 31.5 \\
         \hline
        \multirow[c]{6}{*}{\rotatebox[origin=c]{90}{RREA}} & DInf & 62.8 & 62.8 & 62.8 & 60.4 & 60.3 & 60.4 & 60.5 & 60.5 & 60.5 & 36.7 & 36.7 & 36.7 & 52.1 & 52.1 & 52.1 & 41.6 & 41.6 & 41.6 & 44.8 & 44.8 & 44.8 \\
         & Hun. & \underline{77.8} & 77.7 & 77.7 & \textbf{74.6} & \textbf{74.5} & \textbf{74.6} & \underline{75.0} & 74.9 & \underline{75.0} & 41.9 & 41.7 & 41.8 & 56.5 & 56.2 & 56.4 & 47.7 & 47.6 & 47.6 & \underline{49.7} & 49.5 & 49.6 \\
         & Sink-o & 55.6 & 55.6 & 55.6 & 55.0 & 54.9 & 54.9 & 54.6 & 54.5 & 54.6 & 34.1 & 34.0 & 34.0 & 49.8 & 49.7 & 49.7 & 38.5 & 38.5 & 38.5 & 42.2 & 42.2 & 42.2 \\
         & Sink-d & \textbf{77.8} & \textbf{77.8} & 77.8 & 74.2 & 74.1 & 74.1 & \textbf{75.0} & \textbf{75.0} & \textbf{75.0} & \textbf{42.3} & \textbf{42.2} & \textbf{42.3} & \textbf{56.9} & \textbf{56.7} & \textbf{56.8} & \textbf{48.1} & \textbf{48.0} & \textbf{48.0} & \textbf{49.7} & \textbf{49.7} & \textbf{49.7} \\
         & SMat & 72.2 & 71.8 & 72.0 & 68.3 & 67.7 & 68.0 & 69.2 & 68.6 & 68.9 & 40.3 & 39.8 & 40.1 & 55.8 & 55.0 & 55.4 & 45.7 & 45.3 & 45.5 & 47.6 & 47.2 & 47.4 \\
         & BMat & 72.2 & 71.8 & 72.0 & 68.3 & 67.7 & 68.0 & 69.2 & 68.6 & 68.9 & 40.3 & 39.8 & 40.1 & 55.8 & 55.0 & 55.4 & 45.7 & 45.3 & 45.5 & 47.6 & 47.2 & 47.4 \\
        \bottomrule
    \end{tabular}
\end{table*}
\begin{table}[t]
    \caption{Performance of matching algorithms based on PARIS and GCN-Align for DWY100K dataset.}
    \label{tab:performance_prf1_05_100k}
    \begin{tabular}{|ll|ccc|ccc|}
        \toprule
         &  & \multicolumn{6}{c|}{DWY100K} \\
         \cline{3-8}
         &  & \multicolumn{3}{c|}{DBP\_WD} & \multicolumn{3}{c|}{DBP\_YG} \\
         &  & P & R & F1 & P & R & F1 \\
        \midrule
        \multirow[c]{7}{*}{\rotatebox[origin=c]{90}{PARIS}}
         & Native & \textbf{92.9} & 70.9 & 80.4 & \textbf{95.0} & 46.0 & 62.0 \\
         \cline{3-8}
         & DInf & 85.3 & 71.7 & 77.9 & 80.9 & 46.6 & 59.2 \\
         & Hun. & 91.2 & 73.5 & 81.4 & 89.3 & \textbf{49.0} & \underline{63.2} \\
         & Sink-o & 83.7 & 70.4 & 76.5 & 80.8 & 46.6 & 59.1 \\
         & Sink-d & 87.6 & 73.3 & 79.8 & 86.4 & 48.6 & 62.2 \\
         & SMat & 91.9 & \underline{73.8} & \underline{81.9} & 89.8 & 48.8 & \underline{63.2} \\
         & BMat & 92.1 & \textbf{73.8} & \textbf{81.9} & 89.9 & 48.8 & \textbf{63.2} \\
         \hline
        \multirow[c]{6}{*}{\rotatebox[origin=c]{90}{GCN-Align}}
         & DInf & 40.9 & 40.8 & 40.9 & 55.4 & 55.4 & 55.4 \\
         & Hun. & \textbf{62.1} & \textbf{61.9} & \textbf{62.0} & 73.6 & 73.6 & 73.6 \\
         & Sink-o & 28.6 & 28.4 & 28.5 & 33.5 & 33.5 & 33.5 \\
         & Sink-d & 61.7 & 61.6 & 61.6 & \textbf{74.0} & \textbf{74.0} & \textbf{74.0} \\
         & SMat & 53.4 & 52.8 & 53.1 & 66.8 & 66.6 & 66.7 \\
         & BMat & 53.4 & 52.8 & 53.1 & 66.8 & 66.6 & 66.7 \\
        \bottomrule
    \end{tabular}
\end{table}
Performance of matching algorithms discussed in Section \ref{sec:matching_algorithms}
on the similarity matrix output by PARIS, GCN-Align, RREA are listed in Tables \ref{tab:performance_prf1_05} and \ref{tab:performance_prf1_05_100k} for small-scale (DBP15K, SRPRS) and large-scale (DWY100K) datasets, respectively. 
The performance of PARIS using its own matching mechanism is denoted as Native. In these tables, a higher value means a better performance, and the highest values for each similarity estimation method are shown in bold. Values that are equal to the highest value are also underlined.

It can be seen from these tables that, among similarity estimation methods, PARIS consistently achieves the highest precision for all KG pairs of the three datasets.
This can be explained by the strategy of PARIS to rely on highly discriminative attributes and relations to iteratively estimate and improve the pairwise similarity scores. Additionally, its mechanism to reset matching when the score of already-matched pairs in a previous iteration drops in the following iterations also help to retain highly-probable pairs only. Therefore, PARIS should also be the-method-of-choice when a high precision is expected from the alignment result. For example, its EA result has been used as seeds for supervised algorithms in a neural-symbolic setting \cite{qi_unsupervised_2021}.
Regarding recall and F1-score, PARIS also has the highest values for the three datasets, except for YAGO-related KG pairs. The low recall and low F1-score of PARIS on SRPRS/DBP\_YG and DWY100K/DBP\_YG can be explained by the small number of relations ($|R|$) and the small average number of attributes per entity ($|T^A|/|E|$) in the KG extracted from YAGO, as shown in Table \ref{tab:dataset_statistics}. 
$|R| = 30, 31$ and $|T^A|/|E| = 1.50, 1.18$ for SRPRS/DBP\_YG and DWY100K/DBP\_YG, respectively, which are much smaller than the corresponding values of KG extracted from DBpedia or Wikidata. 
A small value of $|T^A|/|E|$ (i.e., lack of attributes) hinders PARIS to efficiently initialize the similarity score for each entity pair from their attributes. In addition, a small value of $|R|$ means that relations are less unique, making the propagation of similarity scores through them more difficult.
In general, PARIS has very good performance when there is ``enough'' contextual information from the two KGs. It is also robust to different characteristics of KG pairs, such as mono-lingual vs.\ cross-lingual, bijective vs.\ non-bijective.

The performance of two embedding-based methods, GCN-Align and RREA, is in-line with the results recently reported in a recent benchmarking by Zeng et al.\ \cite{zeng_matching_benchmarking_2023}. Since GCN-Align and RREA only use structural information to learn entity embeddings, they are not impacted by the small $|T^A|/|E|$ value in KGs extracted from YAGO. In general, their performance is significant worse than that of PARIS, except for YAGO-related KG pairs. 

\subsection{Performance of matching algorithms}

Matching algorithms behave differently on the similarity matrix output from different similarity estimation methods. For PARIS, since the focus is on reliable matching, its Native matching algorithm has very high precision in exchange for relatively lower recall. This makes room for matching algorithms to improve on the recall.

When combining PARIS with matching algorithms, the best F1-score comes from stable matching (SMat and BMat). SMat and BMat equally improve the F1-score of native PARIS by about 1.9\%, 1.3\%, 1.4\% for DBP15KK, SRPRS, and DWY100K, respectively. Assignment algorithms (DInf, Hun.\, Sink-o, and Sink-d) have lower F1-scores than that of native PARIS for all KG pairs.
There are probably two main reasons for the ``good'' and ``bad'' performance of matching algorithms on the matrix $\mathbf{S}$ output by PARIS. The first is from the 1-to-1 assumption and/or the greedy nature of some algorithms. 1-to-1 assumption helps resolve many-to-1 alignments produced by native PARIS. And greedy finally leaves no entity unmatched.
The second is the difference in the matching objective. For stable matching, it's the ``stability'' of the matching, while for assignment matching, it's the sum of the similarity scores of matched entity pairs. Stable matching might be more suitable with sparse similarity matrix of PARIS in which closely-related entities have very similar similarity scores and the distribution of values is almost bipolar, either 0 or skewed towards 1.
Even though assignment algorithms all decrease the performance of native PARIS, the drop in performance varies and depends on the algorithm and dataset. For example, DInf and Hun.\ have almost the same and better performance than Sink-o and Sink-d when input KGs have high $|R|$ and $|T^A|/|E|$ values in DBP15K. When $|R|$ and $|T^A|/|E|$ values are smaller in SRPRS and DYW100K, Hun.\ performs better than DInf. Among assignment algorithms, Hun.\ and Sink-o have the best and worst matching performance on the matrix $\mathbf{S}$ output by PARIS on the three datasets.

For similarity matrix calculated from the embedding vectors of GCN-Align and RREA, Hun.\ and Sink-d have comparable and best performance, while Sink-o has the worst performance. Hun.\ performs slightly better than Sink-d when input KGs have a high $|R|$ value, such as in DBP15K. When $|R|$ value is small in SRPRS and DYW100K, Hun.\ performs slightly worse than Sink-d.
The F1-score of stable matching algorithms (SMat and BMat) is generally 1\% to 7\% lower than that of Hun.\ and Sink-d. They still, however, have better performance than DInf, which is the ``default'' matching algorithm for embedding-based methods since DInf's F1-score is equivalent to Hit@1 metric when $\theta = 0$ (Equation \ref{eq:matching_with_threshold}).

\subsection{Efficiency and scalability}

Besides performance in terms of precision, recall, and F-1score, the efficiency and scalability of matching algorithms are also important and needs consideration since real-world EA might need to handle millions of entity in each KG \cite{tanon_yago4_2020}. Theoretical time and space complexity of matching algorithms used in our work is already given in Table \ref{tab:algo_characteristics}. 
For efficiency, we observed experimentally that DInf is the most efficient matching algorithm, followed by Sink-o, BMat, SMat, Hun., and Sink-d. The runtime for Sink-o is a multiple of the runtime for DInf, depending on the number of normalization iterations $k$. While BMat and SMat have the same time complexity of $\mathcal{O} (n^2)$, BMat runs faster than SMat due to the generation of preference orders preceding SMat. Hun.\, and Sink-d require much more runtime than the others, especially for the DYW100K dataset, due to their $\mathcal{O}(n^3)$ and $\mathcal{O}(n^3\log n)$ time complexity, respectively.

Even though matching algorithms have the same space complexity of $\mathcal{O}(n^2)$, our adaptation of DInf, Sink-o, SMat, BMat so that they can run on sparse matrix drastically reduces their memory footprint when running on such an input. For Sink-d and Hun., the adopted implementations require a dense input matrix directly.
In our experiments, DInf, Sink-o, SMat, BMat can run on the sparse matrix output by PARIS using a commodity server having only 32 GB memory for all the datasets. All other combinations, either using another similarity estimation method or a different matching algorithm, require a dedicated server of 256 GB memory, especially for the DYW100K dataset.
Thus, the sparsity in $\mathbf{S}$ by PARIS makes some matching algorithms more scalable by requiring much less memory than their theoretical space complexity.

\subsection{Name-based entity alignment}

Matching results using auxiliary information (entity names only) to calculate similarity matrices are listed in Table \ref{tab:performance_name_05}. Even thought entity names are available for SRPRS/DBP\_YG, we don't provide the results since the names of equivalent entity pairs of the two KGs are identical, making name-based matching a trivial problem.
\begin{table*}
    \caption{Performance of matching algorithms based on entity names for DBP15K and SRPRS datasets.}
    \label{tab:performance_name_05}
    \begin{tabular}{|ll|ccc|ccc|ccc|ccc|ccc|ccc|ccc|}
        \toprule
         &  & \multicolumn{9}{c|}{DBP15K} & \multicolumn{6}{c|}{SRPRS} \\
         \cline{3-17}
         &  & \multicolumn{3}{c|}{FR\_EN} & \multicolumn{3}{c|}{JA\_EN} & \multicolumn{3}{c|}{ZH\_EN} & \multicolumn{3}{c|}{EN\_FR} & \multicolumn{3}{c|}{EN\_DE} \\
         &  & P & R & F1 & P & R & F1 & P & R & F1 & P & R & F1 & P & R & F1 \\
        \midrule
        \multirow[c]{6}{*}{\rotatebox[origin=c]{90}{Sorensen-Dice}} & DInf & 68.3 & 85.0 & 75.7 & 62.2 & 74.8 & 67.9 & 54.5 & 62.0 & 58.0 & 93.0 & 89.6 & 91.2 & 93.0 & 88.7 & 90.8 \\
         & Hun. & 75.3 & 88.0 & 81.2 & 71.3 & \textbf{79.1} & 75.0 & \textbf{67.9} & 65.3 & \textbf{66.6} & \textbf{98.3} & \textbf{91.9} & \textbf{95.0} & \textbf{98.5} & \textbf{91.1} & \textbf{94.6} \\
         & Sink-o & 69.4 & 86.5 & 77.0 & 65.1 & 78.4 & 71.1 & 57.3 & 65.2 & 61.0 & 94.3 & 90.9 & 92.6 & 94.4 & 90.1 & 92.2 \\
         & Sink-d & 73.1 & \underline{88.3} & 79.9 & 67.9 & 80.3 & 73.6 & 61.7 & \textbf{66.2} & 63.9 & 96.2 & 91.8 & 94.0 & 96.4 & 90.9 & 93.6 \\
         & SMat & 75.4 & 88.0 & 81.2 & 71.0 & 78.4 & 74.5 & 67.5 & 64.5 & 66.0 & 98.0 & 91.5 & 94.7 & 98.0 & 90.6 & 94.2 \\
         & BMat & \textbf{75.5} & \textbf{88.3} & \textbf{81.4} & \textbf{71.4} & 79.0 & \textbf{75.0} & 67.4 & 64.7 & 66.0 & 98.1 & 91.7 & 94.8 & 98.2 & 90.8 & 94.3 \\
         \hline
        \multirow[c]{6}{*}{\rotatebox[origin=c]{90}{LaBSE}} & DInf & 68.4 & 89.7 & 77.6 & 52.6 & 69.5 & 59.9 & 44.3 & 57.3 & 50.0 & 91.1 & 91.1 & 91.1 & 92.5 & 92.5 & 92.5 \\
         & Hun. & \textbf{72.4} & \textbf{94.9} & \textbf{82.1} & 59.3 & 78.2 & 67.5 & 51.9 & 67.0 & 58.5 & \textbf{95.7} & \textbf{95.7} & \textbf{95.7} & 96.8 & 96.8 & 96.8 \\
         & Sink-o & 67.0 & 87.8 & 76.0 & 51.6 & 68.2 & 58.8 & 43.1 & 55.7 & 48.6 & 91.0 & 91.0 & 91.0 & 92.2 & 92.2 & 92.2 \\
         & Sink-d & \underline{72.4} & \underline{94.9} & \underline{82.1} & \textbf{59.6} & \textbf{78.7} & \textbf{67.8} & \textbf{52.2} & \textbf{67.5} & \textbf{58.9} & 95.6 & 95.6 & 95.6 & \textbf{97.0} & \textbf{97.0} & \textbf{97.0} \\
         & SMat & 71.5 & 93.7 & 81.1 & 57.4 & 75.6 & 65.2 & 49.3 & 63.7 & 55.5 & 94.5 & 94.4 & 94.5 & 96.0 & 95.9 & 95.9 \\
         & BMat & 71.5 & 93.7 & 81.1 & 57.4 & 75.6 & 65.2 & 49.3 & 63.7 & 55.5 & 94.5 & 94.4 & 94.5 & 96.0 & 95.9 & 95.9 \\
         \hline
        \multirow[c]{6}{*}{\rotatebox[origin=c]{90}{fastText}} & DInf & 71.0 & 72.8 & 71.8 & 74.5 & 76.9 & 75.7 & 69.0 & 71.8 & 70.4 & 81.6 & 80.9 & 81.2 & 83.2 & 82.6 & 82.9 \\
         & Hun. & 78.5 & \textbf{76.7} & \textbf{77.6} & 81.7 & \textbf{81.0} & \textbf{81.3} & 75.0 & 74.2 & 74.6 & 89.4 & \textbf{84.9} & \textbf{87.1} & \textbf{90.7} & \textbf{87.3} & \textbf{89.0} \\
         & Sink-o & 69.9 & 71.4 & 70.6 & 73.6 & 75.5 & 74.5 & 67.6 & 70.0 & 68.8 & 81.4 & 80.5 & 81.0 & 82.2 & 81.4 & 81.8 \\
         & Sink-d & 76.4 & 76.6 & 76.5 & 79.6 & 80.6 & 80.1 & 73.5 & \textbf{74.7} & 74.1 & 86.6 & 84.5 & 85.6 & 88.8 & 87.0 & 87.9 \\
         & SMat & \underline{78.8} & 75.6 & 77.2 & \underline{81.9} & 80.0 & 80.9 & \underline{76.3} & 74.0 & \underline{75.2} & \underline{89.8} & 84.3 & 86.9 & 90.6 & 86.5 & 88.5 \\
         & BMat & \textbf{78.8} & 75.6 & 77.2 & \textbf{81.9} & 80.0 & 80.9 & \textbf{76.3} & 74.0 & \textbf{75.2} & \textbf{89.8} & 84.3 & 86.9 & 90.6 & 86.5 & 88.5 \\
        \bottomrule
    \end{tabular}
\end{table*}
It can be observed that name-based matching is better than embedding-based matching (GCN-Align, RREA in Tables \ref{tab:performance_prf1_05}, \ref{tab:performance_prf1_05_100k}). Name-based matching is also better than PARIS for SRPRS dataset. This confirms the utility of discriminative auxiliary information such as entity name in EA. This type of information should be exploited in EA when available.

Among Sorensen-Dice, LaBSE and fastText, Sorensen-Dice generally has a more stable and better performance
than embedding-based methods. This can be explained by the particularity of entity names, it is often short and does not contain rich semantic information. LaBSE and fastText thus cannot ``catch the meaning'' of entity names properly. 
When applying matching algorithms on the matrix $\mathbf{S}$ calculated from Sorensen-Dice, LaBSE and fastText, Hun.\ generally outperforms the others. 

\section{Discussion and Conclusions}

This section firstly summarizes the observations and insights obtained from our experiments in the previous section, then concludes the paper with future research directions.

\textbf{No single silver bullet matching algorithm.} 
When the similarity matrix is sparse, stable matching algorithms have better performance. 
Since SMat and BMat have comparable performance, we hypothesize that the performance of elements of the lattice of stable solutions is also similar.
For dense matrix, better performance is obtained by algorithms designed for the linear assignment problem. 
Hun.\ and Sink-d have comparable and better performance than Sink-o, suggesting that Sinkhorn distance outperforms Sinkhorn operator in solving the matching problem in EA.

\textbf{Conventional method still prevails.} 
When the focus is on precision, the Native matching algorithm of PARIS should be the-method-of-choice. 
Combining PARIS with a stable matching algorithm (SMat or BMat) provides the best F1-score values in most scenarios, except YAGO-related KG pairs due to the lack of contextual information. 
This suggests that deep learning-based models or learning mechanism should integrate key ingredients of PARIS, such as functionality, into their design to get better performance.

\textbf{Embedding-based methods for poor-information KGs.} 
The recall of PARIS drops significantly on YAGO-related KG pairs, indicating the importance of contextual information for pairwise similarity estimation. 
For KGs with poor information, we suggest using an embedding-based similarity estimation method (e.g., RREA) together with an algorithm designed for the linear assignment problem (e.g., Hun.) for better performance.

\textbf{Not all algorithms scale.} 
By taking into account the time complexity and performance of matching algorithms, together with the possibility to reduce memory footprint by working directly on sparse matrix, BMat should be the-method-of-choice among matching algorithms since it consistently have the best performance when working with PARIS. 
Even though SMat has comparable performance with BMat, it has a higher time complexity than BMat.

\textbf{Auxiliary information can make a difference.} 
Entity names are valuable and should be used whenever available. 
Embeddings might not the best way to use names in EA, even though they facilitate the use of names in deep representation learning. 
They are still outperformed by the classic token-based name comparison using Sorensen-Dice coefficient. 
Name information can also be used as an additional attribute for each entity \cite{leone_critical_2022} in order to enrich KG's contextual information, thus improving the final EA performance.
\vspace{0.5em}

\noindent\textbf{Conclusions.} 
We have comprehensively evaluated in this work the performance of matching algorithms on the output of representative similarity estimation methods in EA using three common public datasets. 
We observed that different classes of entity similarity estimation methods may require different matching algorithms to reach the best EA results for each class.
In general, an EA solution composed of PARIS for similarity estimation and a newly proposed BMat for matching gives state-of-the-art results. 
PARIS+BMat is also a scalable solution that can run with large-scale datasets.
Future research should focus on (1) devising novel and scalable matching algorithms specific for the matching problem of EA and, (2) solving EA holistically using graph matching algorithms \cite{fey_deep_2020}.

\section{Acknowledgements}

We would like to thank the authors of the previous benchmarking \cite{zeng_matching_benchmarking_2023} for sharing code and the embeddings of GCN-Align, RREA, and fastText on public datasets used in this work\footnote{\href{https://github.com/DexterZeng/EntMatcher}{https://github.com/DexterZeng/EntMatcher}}.

\bibliographystyle{ACM-Reference-Format}
\bibliography{refs}

\end{document}